\documentclass{article}
\usepackage{amsmath,amssymb,bm}

\usepackage{graphicx}        
\usepackage{multicol}        
\usepackage[bottom]{footmisc}

\title{Introductory lectures on the Effective One Body formalism}
\author{Thibault Damour \\ { \ } \\
Institut des Hautes Etudes Scientifiques, 35 route de Chartres, \\ F-91440 Bures-sur-Yvette, France}
\date{ \ }

\begin{document}

\maketitle

\begin{abstract}
The Effective One Body (EOB) formalism is an analytical approach which aims at providing an accurate description of the motion and radiation of coalescing binary black holes. We present a brief review of the basic elements of this approach.
\end{abstract}

\section{Introduction}\label{sec:1}

A network of ground-based interferometric gravitational wave (GW) detectors (LIGO/VIRGO/GEO/$\ldots$) is currently taking data near its planned sensitivity. Coalescing black hole binaries are among the most promising, and most exciting, GW sources for these detectors. In order to successfully detect GWs from coalescing black hole binaries, and to be able to reliably measure the physical parameters of the source (masses, spins, $\ldots$), it is necessary to know in advance the shape of the GW signals emitted by inspiralling and merging black holes. Indeed, the detection and subsequent data analysis of GW signals is made by using a large bank of {\it templates} that accurately represent the GW waveforms emitted by the source.

\smallskip

Here, we shall introduce the reader to one promising strategy toward having an accurate analytical\footnote{Here we use the adjective `analytical' for methods that solve explicit (analytically given) ordinary differential equations (ODE), even if one uses standard (Runge-Kutta-type) numerical tools to solve them. The important point is that, contrary to $3d$ numerical relativity simulations, numerically solving ODE's is extremely fast, and can therefore be done (possibly even in real time) for a dense sampling of theoretical parameters, such as orbital ($\nu = m_1 \, m_2 / M , \ldots$) or spin ($\hat a_1 = S_1 / Gm_1^2 , \theta_1 , \varphi_1 , \ldots$) parameters.} description of the motion and radiation of binary black holes, which covers all its stages (inspiral, plunge, merger and ring-down): the {\it Effective One Body} approach \cite{Buonanno:1998gg,Buonanno:2000ef,Damour:2000we,Damour:2001tu}. As early as 2000 \cite{Buonanno:2000ef} this method made several quantitative and qualitative predictions concerning the dynamics of the coalescence, and the corresponding GW radiation, notably: (i) a blurred transition from inspiral to a `plunge' that is just a smooth continuation of the inspiral, (ii) a sharp transition, around the merger of the black holes, between a continued inspiral and a ring-down signal, and (iii) estimates of the radiated energy and of the spin of the final black hole. In addition, the effects of the individual spins of the black holes were investigated within the EOB \cite{Damour:2001tu,Buonanno:2005xu} and were shown to lead to a larger energy release for spins parallel to the orbital angular momentum, and to a dimensionless rotation parameter $J/E^2$ always smaller than unity at the end of the inspiral (so that a Kerr black hole can form right after the inspiral phase). All those predictions have been broadly confirmed by the results of the recent numerical simulations performed by several independent groups \cite{gr-qc/0507014,Campanelli:2005dd,Campanelli:2006uy,Baker:2006yw,Baker:2006vn,Baker:2007fb,Gonzalez:2006md,arXiv:0706.0740,Koppitz:2007ev,arXiv:0708.3999,arXiv:0710.3345,arXiv:0710.0158} (for a review of numerical relativity results see \cite{arXiv:0710.1338}). Note that, in spite of the high computer power used in these simulations, the calculation of one sufficiently long waveform (corresponding to specific values of the many continuous parameters describing the two arbitrary masses, the initial spin vectors, and other initial data) takes on the order of two weeks. This is a very strong argument for developing analytical models of waveforms.

\smallskip

Those recent breakthroughs in numerical relativity (NR) open the possibility of comparing in detail the EOB description to NR results. This EOB/NR comparison has been recently initiated in several works \cite{gr-qc/0610122,arXiv:0704.1964,arXiv:0706.3732,arXiv:0704.3550,arXiv:0705.2519,arXiv:0711.2628,arXiv:0712.3003}. The level of analytical/numerical agreement is unprecedented, compared to what has been previously achieved when comparing other types of analytical waveforms to numerical ones. For instance, \cite{arXiv:0711.2628} found that the maximal dephasing between a recent, very accurate NR simulation of 30 GW cycles during late inspiral \cite{arXiv:0710.0158} and EOB could be reduced below $10^{-3}$ GW cycles. The same comparison exhibited also an excellent agreement between the amplitudes of the NR and EOB waveforms. If the reader wishes to put the EOB results in contrast with other (Post-Newtonian or hybrid) approaches he can consult, {\it e.g.}, \cite{arXiv:0704.3764,arXiv:0706.1305,arXiv:0710.0158,arXiv:0710.2335,arXiv:0712.3737,arXiv:0712.3787}.

\smallskip

Before reviewing some of the technical aspects of the EOB method, let us indicate some of the historical roots of this method. First, we note that the EOB approach comprises three, rather separate, ingredients:
\begin{itemize}
\item[(1)] a description of the conservative (Hamiltonian) part of the dynamics of two black holes;
\item[(2)] an expression for the radiation-reaction part of the dynamics;
\item[(3)] a description of the GW waveform emitted by a coalescing binary system.
\end{itemize}

\medskip

For each one of these ingredients, the essential inputs that are used in EOB developments are high-order post-Newtonian (PN) expanded results which have been obtained by many years of work, by many researchers (see references below). However, one of the key ideas in the EOB philosophy is to avoid using PN results in their original `Taylor-expanded' form ({\it i.e.} $c_0 + c_1 \, v + c_2 \, v^2 + c_3 \, v^3 + \cdots + c_n \, v^n)$, but to use them instead in some {\it resummed} form ({\it i.e.} some non-polynomial function of $v$, defined so as to incorporate some of the expected non-perturbative features of the exact result). The basic ideas and techniques for resumming each ingredient of the EOB are different and have different historical roots. Concerning the ingredient (1), {\it i.e.} the EOB Hamiltonian, it was inspired by an approach to electromagnetically interacting quantum two-body systems introduced by Br\'ezin, Itzykson and Zinn-Justin \cite{Brezin:1970zr}. The resummation of the ingredient (2), {\it i.e.} the EOB radiation-reaction force ${\mathcal F}$, was inspired by the Pad\'e resummation of the flux function introduced by Damour, Iyer and Sathyaprakash \cite{Damour:1997ub}. As for the ingredient (3), {\it i.e.} the EOB description of the waveform emitted by a coalescing black hole binary, it was mainly inspired by the work of Davis, Ruffini and Tiomno \cite{Davis:1972ud} which discovered the transition between the plunge signal and a ringing tail when a particle falls into a black hole. Additional motivation for the EOB treatment of the transition from plunge to ring-down came from work on the, so-called, `close limit approximation' \cite{Price:1994pm}.

\smallskip

Let us finally note that the EOB approach has been recently improved by following a methodology consisting of studying, element by element, the physics behind each feature of the waveform, and on systematically comparing various EOB-based waveforms with `exact' waveforms obtained by NR approaches. Among these `exact' NR waveforms, it has been useful to consider the small-mass-ratio limit $\nu \equiv m_1 \, m_2 / (m_1 + m_2)^2 \ll 1$, in which one can use the well controllable `laboratory' of numerical simulations of test particles (with an added radiation-reaction force) moving in black hole backgrounds \cite{arXiv:0705.2519}.

\section{Motion and radiation of binary black holes: post-Newtonian expanded results}\label{sec:2}

Before discussing the various resummation techniques used in the EOB approach, let us briefly recall the `Taylor-expanded' results that have been obtained by pushing to high accuracies the post-Newtonian (PN) methods.

\smallskip

Concerning the orbital dynamics of compact binaries, we recall that the 2.5PN-accurate\footnote{As usual `$n$-PN accuracy' means that a result has been derived up to (and including) terms which are $\sim (v/c)^{2n} \sim (GM/c^2 r)^n$ fractionally smaller than the leading contribution.} equations of motion have been derived in the 1980's \cite{Damour:1981bh,Damour:1982wm,Schafer:1986rd,kopejkin1985}. Pushing the accuracy of the equations of motion to the 3PN ($\sim (v/c)^6$) level proved to be a non-trivial task. At first, the representation of black holes by delta-function sources and the use of the (non diffeomorphism invariant) Hadamard regularization method led to ambiguities in the computation of the badly divergent integrals that enter the 3PN equations of motion \cite{gr-qc/9712075,gr-qc/0007051}. This problem was solved by using the (diffeomorphism invariant) {\it dimensional regularization} method ({\it i.e.} analytic continuation in the dimension of space $d$) which allowed one to complete the determination of the 3PN-level equations of motion \cite{gr-qc/0105038,gr-qc/0311052}. They have also been derived by an Einstein-Infeld-Hoffmann-type surface-integral approach \cite{IF03}. The 3.5PN terms in the equations of motion are also known \cite{gr-qc/0201001,KonigsdorfferFayeSchafer03,gr-qc/0412018}.

\smallskip

Concerning the emission of gravitational radiation, two different {\it gravitational-wave generation formalisms} have been developed up to a high PN accuracy: (i) the Blanchet-Damour-Iyer formalism \cite{BD86,BD89,DI91a,DI91b,BD92,gr-qc/9501030,gr-qc/9710038} combines a multipolar post-Minkowskian (MPM) expansion in the exterior zone with a post-Newtonian expansion in the near zone; while (ii) the Will-Wiseman-Pati formalism \cite{gr-qc/9608012,gr-qc/9910057,gr-qc/0007087,gr-qc/0201001} uses a direct integration of the relaxed Einstein equations. These formalisms were used to compute increasingly accurate estimates of the gravitational waveforms emitted by inspiralling binaries. These estimates include both normal, near-zone generated post-Newtonian effects (at the 1PN \cite{BD89}, 2PN \cite{gr-qc/9501027,gr-qc/9501029,gr-qc/9608012}, and 3PN \cite{gr-qc/0105098,gr-qc/0409094} levels), and more subtle, wave-zone generated (linear and non-linear) `tail effects' \cite{BD92,W93,BS93,gr-qc/9710038}. However, technical problems arose at the 3PN level. The representation of black holes by `delta-function' sources causes the appearance of dangerously divergent integrals in the 3PN multipole moments. The use of Hadamard (partie finie) regularization did not allow one to unambiguously compute the needed 3PN-accurate quadrupole moment. Only the use of the (formally) diffeomorphism-invariant {\it dimensional regularization} method allowed one to complete the 3PN-level gravitational-radiation formalism \cite{gr-qc/0503044}.

\smallskip

The works mentioned in this Section (see \cite{Blivingreview} for a detailed account and more references) finally lead to PN-expanded results for the motion and radiation of binary black holes. For instance, the equations of motion are given in the form ($a=1,2$; $i=1,2,3$)
\begin{equation}
\label{eq6.1}
\frac{d^2 z_a^i}{dt^2} = A_a^{i \, {\rm cons}} + A_a^{iRR} \, ,
\end{equation}
where 
\begin{equation}
\label{eq6.2}
A^{\rm cons} = A_0 + c^{-2} A_2 + c^{-4} A_4 + c^{-6} A_6 \, ,
\end{equation}
denotes the `conservative' 3PN-accurate terms, while
\begin{equation}
\label{eq6.3}
A^{RR} = c^{-5} A_5 + c^{-7} A_7 \, ,
\end{equation}
denotes the time-asymmetric contibutions, linked to `radiation reaction'.

\smallskip

On the other hand, if we consider for simplicity the inspiralling motion of a quasi-circular binary system, the essential quantity describing the emitted gravitational waveform is the {\it phase} $\phi$ of the quadrupolar gravitational wave amplitude $h(t) \simeq a(t) \cos (\phi (t) + \delta)$. PN theory allows one to derive several different functional expressions for the gravitational wave phase $\phi$, as a function either of time or of the instantaneous frequency. For instance, as a function of time, $\phi$ admits the following explicit expansion in powers of $\theta \equiv \nu c^3 (t_c - t) / 5GM$ (where $t_c$ denotes a formal `time of coalescence', $M \equiv m_1 + m_2$ and $\nu \equiv m_1 \, m_2 / M^2$)
\begin{equation}
\label{eq6.4}
\phi (t) = \phi_c - \nu^{-1} \, \theta^{5/8} \left( 1 + \sum_{n=2}^7 (a_n + a'_n \, \ln \, \theta) \, \theta^{-n/8} \right) , 
\end{equation}
with some numerical coefficients $a_n , a'_n$ which depend only on the dimensionless (symmetric) mass ratio $\nu \equiv m_1 \, m_2 / M^2$. The derivation of the 3.5PN-accurate expansion (\ref{eq6.4}) uses both the 3PN-accurate conservative acceleration (\ref{eq6.2}) and a 3.5PN extension of the (fractionally) 1PN-accurate radiation reaction acceleration (\ref{eq6.3}) obtained by assuming a balance between the energy of the binary system and the gravitational-wave energy flux at infinity (see, {\it e.g.}, \cite{Blivingreview}).

\section{Conservative dynamics of binary black holes: the Effective One Body approach}\label{sec:3}

The PN-expanded results briefly reviewed in the previous Section are expected to yield accurate descriptions of the motion and radiation of binary black holes only during their {\it early inspiralling} stage, {\it i.e.} as long as the PN expansion parameter $\gamma_e = GM/c^2 R$ (where $R$ is the distance between the two black holes) stays significantly smaller than the value $\sim \frac{1}{6}$ where the orbital motion is expected to become dynamically unstable (`last stable circular orbit' and beginning of a `plunge' leading to the merger of the two black holes). One needs a better description of the motion and radiation to describe the {\it late inspiral} (say $\gamma_e \gtrsim \frac{1}{12}$), as well as the subsequent {\it plunge} and {\it merger}. One possible strategy for having a complete description of the motion and radiation of binary black holes, covering all the stages (inspiral, plunge, merger, ring-down), would then be to try to `stitch together' PN-expanded analytical results describing the early inspiral phase with $3d$ numerical results describing the end of the inspiral, the plunge, the merger and the ring-down of the final black hole, see, {\it e.g.}, Refs.~\cite{gr-qc/0612117,arXiv:0704.1964}.

\smallskip

However, we wish to argue that the EOB approach makes a better use of all the analytical information contained in the PN-expanded results (\ref{eq6.1})-(\ref{eq6.3}). The basic claim (first made in \cite{Buonanno:1998gg,Buonanno:2000ef}) is that the use of suitable {\it resummation methods} should allow one to describe, by analytical tools, a {\it sufficiently accurate} approximation of the {\it entire waveform}, from inspiral to ring-down, including the non-perturbative plunge and merger phases. To reach such a goal, one needs to make use of several tools: (i) resummation methods, (ii) exploitation of the flexibility of analytical approaches, (iii) extraction of the non-perturbative information contained in various numerical simulations, (iv) qualitative understanding of the basic physical features which determine the waveform.

\smallskip

Let us start by discussing the first tool used in the EOB approach: the systematic use of resummation methods. Two such methods have been employed (and combined), and some evidence has been given that they do significantly improve the convergence properties of PN expansions. The first method is the use of {\it Pad\'e approximants}. It has been shown in Ref.~\cite{Damour:1997ub} that near-diagonal Pad\'e approximants of the radiation reaction force\footnote{We henceforth denote by ${\mathcal F}$ the {\it Hamiltonian} version of the radiation reaction term $A^{RR}$, Eq.~(\ref{eq6.3}), in the (PN-expanded) equations of motion. It can be heuristically computed up to (absolute) 5.5PN \cite{gr-qc/0105098, gr-qc/0406012,gr-qc/0503044} and even 6PN \cite{gr-qc/0105099} order by assuming that the energy radiated in gravitational waves at infinity is balanced by a loss of the dynamical energy of the binary system.} ${\mathcal F}$ seemed to provide a good representation of ${\mathcal F}$ down to the last stable orbit (which is expected to occur when $R \sim 6GM/c^2$, {\it i.e.} when $\gamma_e \simeq \frac{1}{6}$). The second method is a novel approach to the dynamics of compact binaries, which constitutes the core of the Effective One Body (EOB) method. 

\smallskip

For simplicity of exposition, let us first explain the EOB method at the 2PN level. The starting point of the method is the 2PN-accurate Hamiltonian describing (in Arnowitt-Deser-Misner-type coordinates) the conservative, or time symmetric, part of the equations of motion (\ref{eq6.1}) ({\it i.e.} the truncation $A^{\rm cons} = A_0 + c^{-2} A_2 + c^{-4} A_4$ of Eq.~(\ref{eq6.2})) say $H_{\rm 2PN} ({\bm q}_1 - {\bm q}_2 , {\bm p}_1 , {\bm p}_2)$. By going to the center of mass of the system $({\bm p}_1 + {\bm p}_2 = 0)$, one obtains a PN-expanded Hamiltonian describing the {\it relative motion}, ${\bm q} = {\bm q}_1 - {\bm q}_2$, ${\bm p} = {\bm p}_1 = - {\bm p}_2$:
\begin{equation}
\label{eq7.1}
H_{\rm 2PN}^{\rm relative} ({\bm q} , {\bm p}) = H_0 ({\bm q} , {\bm p}) + \frac{1}{c^2} \, H_2 ({\bm q} , {\bm p}) + \frac{1}{c^4} \, H_4 ({\bm q} , {\bm p}) \, ,
\end{equation}
where $H_0 ({\bm q} , {\bm p}) = \frac{1}{2\mu} \, {\bm p}^2 + \frac{GM\mu}{\vert {\bm q} \vert}$ (with $M \equiv m_1 + m_2$ and $\mu = m_1 \, m_2 / M$) corresponds to the Newtonian approximation to the relative motion, while $H_2$ describes 1PN corrections and $H_4$ 2PN ones. It is well known that, at the Newtonian approximation, $H_0 ({\bm q} , {\bm p})$ can be thought of as describing a `test particle' of mass $\mu$ orbiting around an `external mass' $GM$. The EOB approach is a {\it general relativistic generalization} of this fact. It consists in looking for an `external spacetime geometry' $g_{\mu\nu}^{\rm ext} (x^{\lambda} ; GM)$ such that the geodesic dynamics of a `test particle' of mass $\mu$ within $g_{\mu\nu}^{\rm ext} (x^{\lambda} , GM)$ is {\it equivalent}  (when expanded in powers of $1/c^2$) to the original, relative PN-expanded dynamics (\ref{eq7.1}). 

\smallskip 

Let us explain the idea, proposed in \cite{Buonanno:1998gg}, for establishing a `dictionary' between the real relative-motion dynamics, (\ref{eq7.1}), and the dynamics of an `effective' particle of mass $\mu$ moving in $g_{\mu\nu}^{\rm ext} (x^{\lambda} , GM)$. The idea consists in `thinking quantum mechanically'\footnote{This is related to an idea emphasized many times by John Archibald Wheeler: quantum mechanics can often help us in going to the essence of classical mechanics.}. Instead of thinking in terms of a classical Hamiltonian, $H({\bm q}, {\bm p})$ (such as $H_{\rm 2PN}^{\rm relative}$, Eq.~(\ref{eq7.1})), and of its classical bound orbits, we can think in terms of the quantized energy levels $E(n,\ell)$ of the quantum bound states of the Hamiltonian operator $H (\hat{\bm q} , \hat{\bm p})$. These energy levels will depend on two (integer valued) quantum numbers $n$ and $\ell$. Here (for a spherically symmetric interaction, as appropriate to $H^{\rm relative}$), $\ell$ parametrizes the total orbital angular momentum (${\bm L}^2 = \ell (\ell + 1) \, \hbar^2$), while $n$ represents the `principal quantum number' $n = \ell + n_r + 1$, where $n_r$ (the `radial quantum number') denotes the number of nodes in the radial wave function. The third `magnetic quantum number' $m$ (with $-\ell \leq m \leq \ell$) does not enter the energy levels because of the spherical symmetry of the two-body interaction (in the center of of mass frame). For instance, a non-relativistic Coulomb (or Newton!) interaction
\begin{equation}
\label{eqn1}
H_0 = \frac{1}{2\mu} \, {\bm p}^2 + \frac{GM\mu}{\vert {\bm q} \vert}
\end{equation}
gives rise to the well-known result
\begin{equation}
\label{eqn2}
E_0 (n,\ell) = - \frac{1}{2} \, \mu \left(\frac{GM\mu}{n \, \hbar} \right)^2 \, ,
\end{equation}
which depends only on $n$ (this is the famous Coulomb degeneracy). When considering the PN corrections to $H_0$, as in Eq.~(\ref{eq7.1}), one gets a more complicated expression of the form
\begin{eqnarray}
\label{eqn3}
E_{\rm 2PN}^{\rm relative} (n,\ell) = - \frac{1}{2} \, \mu \ \frac{\alpha^2}{n^2} &&\!\!\!\!\!\!\!\!\!\!\biggl[ 1 + \frac{\alpha^2}{c^2} \left( \frac{c_{11}}{n\ell} + \frac{c_{20}}{n^2} \right) \nonumber \\
&&\!\!+ \frac{\alpha^4}{c^4} \left( \frac{c_{13}}{n\ell^3} + \frac{c_{22}}{n^2 \ell^2} + \frac{c_{31}}{n^3 \ell} + \frac{c_{40}}{n^4} \right)\biggl] \, ,
\end{eqnarray}
where we have set $\alpha \equiv GM\mu / \hbar = G \, m_1 \, m_2 / \hbar$, and where we consider, for simplicity, the (quasi-classical) limit where $n$ and $\ell$ are large numbers. The 2PN-accurate result (\ref{eqn3}) had been derived by Damour and Sch\"afer \cite{Damour:1988mr} as early as 1988. The dimensionless coefficients $c_{pq}$ are functions of the symmetric mass ratio $\nu \equiv \mu / M$, for instance $c_{40} = \frac{1}{8} (145 - 15\nu + \nu^2)$. In classical mechanics ({\it i.e.} for large $n$ and $\ell$), it is called the `Delaunay Hamiltonian', {\it i.e.} the Hamiltonian expressed in terms of the {\it action variables}\footnote{We consider, for simplicity, `equatorial' motions with $m=\ell$, {\it i.e.}, classically, $\theta = \frac{\pi}{2}$.} $J = \ell \hbar = \frac{1}{2\pi} \oint p_{\varphi} \,d\varphi$, and $N = n \hbar = I_r + J$, with $I_r = \frac{1}{2\pi} \oint p_r \, dr$.

\smallskip

The energy levels (\ref{eqn3}) encode, in a gauge-invariant way, the 2PN-accurate relative dynamics of a `real' binary. Let us now consider an auxiliary problem: the `effective' dynamics of one body, of mass $\mu$, following a geodesic in some `external' (spherically symmetric) metric\footnote{It is convenient to write the `external metric' in Schwarzschild-like coordinates. Note that the external radial coordinate $R$ differs from the two-body ADM-coordinate relative distance $R^{\rm ADM} = \vert {\bm q} \vert$. The transformation between the two coordinate systems has been determined in Refs.~\cite{Buonanno:1998gg,Damour:2000we}.}
\begin{equation}
\label{eq7.2}
g_{\mu\nu}^{\rm ext} \, dx^{\mu} \, dx^{\nu} = - A(R) \, c^2 \, d T^2 + B(R) \, d R^2 + R^2 (d\theta^2 + \sin^2 \theta \, d \varphi^2) \, .
\end{equation}
Here, the {\it a priori unknown} metric functions $A(R)$ and $B(R)$ will be constructed in the form of an expansion in $GM/c^2 R$:
\begin{eqnarray}
\label{eqn4}
A(R) &= &1 + a_1 \, \frac{GM}{c^2 R} + a_2 \left( \frac{GM}{c^2 R} \right)^2 + a_3 \left( \frac{GM}{c^2 R} \right)^3 + \cdots \, ; \nonumber \\
B(R) &= &1 + b_1 \, \frac{GM}{c^2 R} + b_2 \left( \frac{GM}{c^2 R} \right)^2 + \cdots \, ,
\end{eqnarray}
where the dimensionless coefficients $a_n , b_n$ depend on $\nu$. From the Newtonian limit, it is clear that we should set $a_1 = -2$. By solving (by separation of variables) the `effective' Hamilton-Jacobi equation
$$
g_{\rm eff}^{\mu\nu} \, \frac{\partial S_{\rm eff}}{\partial x^{\mu}} \, \frac{\partial S_{\rm eff}}{\partial x^{\nu}} + \mu^2 c^2 = 0 \, ,
$$
\begin{equation}
\label{eqn5}
S_{\rm eff} = - {\mathcal E}_{\rm eff} \, t + J_{\rm eff} \, \varphi + S_{\rm eff} (R) \, ,
\end{equation}
one can straightforwardly compute (in the quasi-classical, large quantum numbers limit) the Delaunay Hamiltonian ${\mathcal E}_{\rm eff} (N_{\rm eff} , J_{\rm eff})$, with $N_{\rm eff} = n_{\rm eff} \, \hbar$, $J_{\rm eff} = \ell_{\rm eff} \, \hbar$ (where $N_{\rm eff} = J_{\rm eff} + I_R^{\rm eff}$, with $I_R^{\rm eff} = \frac{1}{2\pi} \oint p_R^{\rm eff} \, dR$, $P_R^{\rm eff} = \partial S_{\rm eff} (R) / dR$). This  yields a result of the form
\begin{eqnarray}
\label{eqn6}
{\mathcal E}_{\rm eff} (n_{\rm eff},\ell_{\rm eff}) = \mu c^2 - \frac{1}{2} \, \mu \ \frac{\alpha^2}{n_{\rm eff}^2} &&\!\!\!\!\!\!\!\!\!\!\biggl[ 1 + \frac{\alpha^2}{c^2} \left( \frac{c_{11}^{\rm eff}}{n_{\rm eff} \ell_{\rm eff}} + \frac{c_{20}^{\rm eff}}{n_{\rm eff}^2} \right) \\
&&\!\!+ \frac{\alpha^4}{c^4} \left( \frac{c_{13}^{\rm eff}}{n_{\rm eff} \ell_{\rm eff}^3} + \frac{c_{22}^{\rm eff}}{n_{\rm eff}^2 \ell_{\rm eff}^2} + \frac{c_{31}^{\rm eff}}{n_{\rm eff}^3 \ell_{\rm eff}} + \frac{c_{40}^{\rm eff}}{n_{\rm eff}^4} \right)\biggl] \, , \nonumber 
\end{eqnarray}
where the dimensionless coefficients $c_{pq}^{\rm eff}$ are now functions of the unknown coefficients $a_n , b_n$ entering the looked for `external' metric coefficients (\ref{eqn4}).

\smallskip

At this stage, one needs (as in the famous AdS/CFT correspondence) to define a `dictionary' between the real (relative) two-body dynamics, summarized in Eq.~(\ref{eqn3}), and the effective one-body one, summarized in Eq.~(\ref{eqn6}). As, on both sides, quantum mechanics tells us that the action variables are quantized in integers ($N_{\rm real} = n \hbar$, $N_{\rm eff} = n_{\rm eff} \hbar$, etc.) it is most natural to identify $n=n_{\rm eff}$ and $\ell = \ell_{\rm eff}$. One then still needs a rule for relating the two different energies $E_{\rm real}^{\rm relative}$ and ${\mathcal E}_{\rm eff}$. Ref.~\cite{Buonanno:1998gg} proposed to look for a general map between the real energy levels and the effective ones (which, as seen when comparing (\ref{eqn3}) and (\ref{eqn6}), cannot be directly identified because they do not include the same rest-mass contribution\footnote{Indeed $E_{\rm real}^{\rm total} = Mc^2 + E_{\rm real}^{\rm relative} = Mc^2 + \mbox{Newtonian terms} + {\rm 1PN} / c^2 + \cdots$, while ${\mathcal E}_{\rm effective} = \mu c^2 + N + {\rm 1PN} / c^2 + \cdots$.}), namely
\begin{equation}
\label{eqn7}
\frac{{\mathcal E}_{\rm eff}}{\mu c^2} - 1 = f \left( \frac{E_{\rm real}^{\rm relative}}{\mu c^2} \right) = \frac{E_{\rm real}^{\rm relative}}{\mu c^2} \left( 1 + \alpha_1 \, \frac{E_{\rm real}^{\rm relative}}{\mu c^2} + \alpha_2 \left( \frac{E_{\rm real}^{\rm relative}}{\mu c^2} \right)^2 + \cdots \right) \, .
\end{equation}
The `correspondence' between the real and effective energy levels is illustrated in Fig.~1.

\begin{figure}[h]
\centering
\includegraphics[height=8cm]{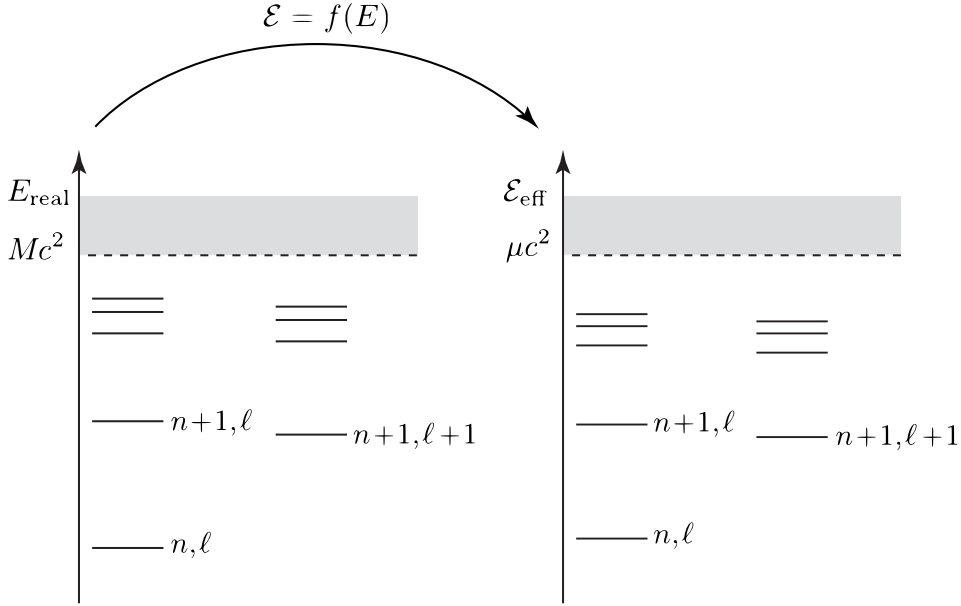}
\caption{Sketch of the correspondence between the quantized energy levels of the real and effective conservative dynamics. $n$ denotes the `principal quantum number' ($n = n_r + \ell + 1$, with $n_r = 0,1,\ldots$ denoting the number of nodes in the radial function), while $\ell$ denotes the (relative) orbital angular momentum $({\bm L}^2 = \ell (\ell + 1) \, \hbar^2)$. Though the EOB method is purely classical, it is conceptually useful to think in terms of the underlying (Bohr-Sommerfeld) quantization conditions of the action variables $I_R$ and $J$ to motivate the identification between $n$ and $\ell$ in the two dynamics.}
\label{fig:1}       
\end{figure}

Finally, identifying ${\mathcal E}_{\rm eff} (n,\ell) / \mu c^2$ to $f (E_{\rm real}^{\rm relative} / \mu c^2)$ yields six equations, relating the six coefficients $c_{pq}^{\rm eff} (a_2 , a_3 ; b_1 , b_2)$ to the six $c_{pq} (\nu)$ and to the two energy coefficients $\alpha_1$ and $\alpha_2$. It is natural to set $b_1 = + 2$ (so that the linearized effective metric coincides with the linearized Schwarzschild metric with mass $M = m_1 + m_2$). One then finds that there exists a {\it unique} solution for the remaining five unknown coefficients $a_2 , a_3 , b_2 , \alpha_1$ and $\alpha_2$. This solution is very simple:
\begin{equation}
\label{eqn8}
a_2 = 0  \, , \quad a_3 = 2 \nu \, , \quad b_2 = 4 - 6 \nu \, , \quad \alpha_1 = \frac{\nu}{2} \, , \quad \alpha_2 = 0 \, .
\end{equation}
Note, in particular, that the map between the two energies is simply
\begin{equation}
\label{eqn9}
\frac{{\mathcal E}_{\rm eff}}{\mu c^2} = 1 + \frac{E_{\rm real}^{\rm relative}}{\mu c^2} \left( 1 + \frac{\nu}{2} \, \frac{E_{\rm real}^{\rm relative}}{\mu c^2} \right) = \frac{s - m_1^2 \, c^4 - m_2^2 \, c^4}{2 \, m_1 \, m_2 \, c^4}
\end{equation}
where $s = ({\mathcal E}_{\rm real}^{\rm tot})^2 \equiv (M c^2 + E_{\rm real}^{\rm relative})^2$ is Mandelstam's invariant. Note also that, at 2PN accuracy, the crucial `$g_{00}^{\rm ext}$' metric coefficient $A(R)$ (which fully encodes the energetics of circular orbits) is given by the remarkably simple PN expansion
\begin{equation}
\label{eq7.3}
A_{\rm 2PN} (R) = 1-2u + 2 \, \nu \, u^3 \, ,
\end{equation}
where $u \equiv GM/c^2 R$ and $\nu \equiv \mu / M \equiv m_1 \, m_2 / (m_1 + m_2)^2$.

\smallskip

The dimensionless parameter $\nu \equiv \mu / M$ varies between $0$ (in the test mass limit $m_1 \ll m_2$) and $\frac{1}{4}$ (in the equal-mass case $m_1 = m_2$). When $\nu \to 0$, Eq.~(\ref{eq7.3}) yields back, as expected, the well-known Schwarzschild time-time metric coefficient $- g_{00}^{\rm Schw} = 1 - 2u = 1 - 2GM / c^2 R$. One therefore sees in Eq.~(\ref{eq7.3}) the r\^ole of $\nu$ as a deformation parameter connecting a well-known test-mass result to a non trivial and new 2PN result. It is also to be noted that the 1PN EOB result $A_{\rm 1PN} (R) = 1-2u$ happens to be $\nu$-independent, and therefore identical to $A^{\rm Schw} = 1-2u$. This is remarkable in view of the many non-trivial $\nu$-dependent terms in the 1PN relative dynamics. The physically real 1PN $\nu$-dependence happens to be fully encoded in the function $f(E)$ mapping the two energy spectra given in Eq.~(\ref{eqn9}) above.

\smallskip

Let us emphasize the remarkable simplicity of the 2PN result (\ref{eq7.3}). The 2PN Hamiltonian (\ref{eq7.1}) contains eleven rather complicated $\nu$-dependent terms. After transformation to the EOB format, the dynamical information contained in these eleven coefficients gets {\it compactified} into the very simple additional contribution $+ \, 2 \, \nu \, u^3$ in $A(R)$, together with an equally simple contribution in the radial metric coefficient: $(A(R) \, B(R))_{\rm 2PN} = 1 - 6 \, \nu \, u^2$. This compactification process is even more drastic when one goes to the next (conservative) post-Newtonian order: the 3PN level, i.e. additional terms of order ${\mathcal O} (1/c^6)$ in the Hamiltonian (\ref{eq7.1}). As mentioned above, the complete obtention of the 3PN dynamics has represented quite a theoretical challenge and the final, resulting Hamiltonian is quite complicated. Even after going to the center of mass frame, the 3PN additional contribution $\frac{1}{c^6} \, H_6 ({\bm q} , {\bm p})$ to Eq.~(\ref{eq7.1}) introduces eleven new complicated $\nu$-dependent coefficients. After transformation to the EOB format \cite{Damour:2000we}, these eleven new coefficients get `compactified' into only {\it three} additional terms: (i) an additional contribution to $A(R)$, (ii) an additional contribution to $B(R)$, and (iii) a ${\mathcal O} ({\bm p}^4)$ modification of the `external' geodesic Hamiltonian. For instance, the crucial 3PN $g_{00}^{\rm ext}$ metric coefficient becomes
\begin{equation}
\label{eq7.5}
A_{\rm 3PN} (R) = 1-2u + 2 \, \nu \, u^3 + a_4 \, \nu \, u^4 \, ,
\end{equation}
where
\begin{equation}
\label{eq7.6}
a_4 = \frac{94}{3} - \frac{41}{32} \, \pi^2 \simeq 18.6879027 \, .
\end{equation}
Remarkably, it is found that the very simple 2PN energy map Eq.~(\ref{eqn9}) does not need to be modified at the 3PN level.

\smallskip

The fact that the 3PN coefficient $a_4$ in the crucial `effective radial potential' $A_{\rm 3PN} (R)$, Eq.~(\ref{eq7.5}), is rather large and positive indicates that the $\nu$-dependent nonlinear gravitational effects lead, for comparable masses $(\nu \sim \frac{1}{4}$), to a last stable (circular) orbit (LSO) which has a higher frequency and a larger binding energy than what a naive scaling from the test-particle limit $(\nu \to 0)$ would suggest. Actually, the PN-expanded form (\ref{eq7.5}) of $A_{\rm 3PN} (R)$ does not seem to be a good representation of the (unknown) exact function $A_{\rm EOB} (R)$ when the (Schwarzschild-like) relative coordinate $R$ becomes smaller than about $6 GM / c^2$ (which is the radius of the LSO in the test-mass limit). It was therefore suggested \cite{Damour:2000we} to further resum\footnote{The PN-expanded EOB building blocks $A(R) , B(R) , \ldots$ already represent a {\it resummation} of the PN dynamics in the sense that they have compactified the many terms of the original PN-expanded Hamiltonian within a very concise format. But one should not refrain to further resum the EOB building blocks themselves, if this is physically motivated.} $A_{\rm 3PN} (R)$ by replacing it by a suitable Pad\'e $(P)$ approximant. For instance, the replacement of $A_{\rm 3PN} (R)$ by
\begin{equation}
\label{eq7.7}
A_3^1 (R) \equiv P_3^1 [A_{\rm 3PN} (R)] = \frac{1+n_1 u}{1+d_1 u + d_2 u^2 + d_3 u^3}
\end{equation}
ensures that the $\nu = \frac{1}{4}$ case is smoothly connected with the $\nu = 0$ limit.

\smallskip

The use of (\ref{eq7.7}) was suggested before one had any (reliable) non-perturbative information on the binding of close black hole binaries. Later, a comparison with some `waveless' numerical simulations of circular black hole binaries \cite{gr-qc/0204011} has given some evidence that (\ref{eq7.7}) is physically adequate. In Refs.~\cite{Damour:2001tu,gr-qc/0204011} it was also emphasized that, in principle, the comparison between numerical data and EOB-based predictions should allow one to determine the effect of the unknown higher PN contributions to Eq.~(\ref{eq7.5}). For instance, one can add a 4PN-like term $+ \, a_5 \, \nu \, u^5$ in Eq.~(\ref{eq7.5}), and then Pad\'e the resulting radial function, say $A_4^1 = P_4^1 [A_{\rm 3PN} + a_5 \, \nu \, u^5]$. Comparing the predictions of $A_4^1 [a_5]$ to numerical data might then determine what is the physically preferred `effective' value of the unknown coefficient $a_5$. This is an example of the useful {\it flexibility} \cite{gr-qc/0211041} of analytical approaches: the fact that one can tap numerically-based, non-perturbative information to improve the EOB approach.

\section{Description of radiation-reaction effects in the Effective One Body approach}\label{sec:4}

In the previous Section we have described how the EOB method encodes the conservative part of the relative orbital dynamics into the dynamics of an 'effective' particle. Let us now briefly discuss how to complete the EOB dynamics by defining a {\it resummed} expression describing radiation reaction effects. One is interested in circularized binaries, which have lost their initial eccentricity under the influence of radiation reaction. For such systems, it is enough to include a radiation reaction force in the $p_{\varphi}$ equation of motion. More precisely, we are using phase space variables $r , p_r , \varphi , p_{\varphi}$ associated to polar coordinates (in the equatorial plane $\theta = \frac{\pi}{2}$). Actually it is convenient to replace the radial momentum $p_r$ by the momentum conjugate to the `tortoise' radial coordinate $R_* = \int dR (B/A)^{1/2}$, {\it i.e.} $P_{R_*} = (A/B)^{1/2} \, P_R$. The real EOB Hamiltonian is obtained by first solving Eq.~(\ref{eqn9}) to get $E_{\rm real}^{\rm total} = \sqrt s$ in terms of ${\mathcal E}_{\rm eff}$, and then by solving the effective Hamiltonian-Jacobi equation\footnote{Completed by the ${\mathcal O} ({\bm p}^4)$ terms that must be introduced at 3PN.} to get ${\mathcal E}_{\rm eff}$ in terms of the effective phase space coordinates ${\bm q}_{\rm eff}$ and ${\bm p}_{\rm eff}$. The result is given by two nested square roots (we henceforth set $c=1$):
\begin{equation}
\label{eqn10}
\hat H_{\rm EOB} (r,p_{r_*} , \varphi) = \frac{H_{\rm EOB}^{\rm real}}{M} = \sqrt{1 + 2 \nu \, (\hat H_{\rm eff} - 1)} \, ,
\end{equation}
where
\begin{equation}
\label{eqn11}
\hat H_{\rm eff} = \sqrt{p_{r_*}^2 + A(r) \left( 1 + \frac{p_{\varphi}^2}{r^2} + z_3 \, \frac{p_{r_*}^4}{r^2} \right)} \, ,
\end{equation}
with $z_3 = 2\nu \, (4-3\nu)$. Here, we are using suitably rescaled dimensionless (effective) variables: $r = R/GM$, $p_{r_*} = P_{R_*} / \mu$, $p_{\varphi} = P_{\varphi} / \mu \, GM$, as well as a rescaled time $t = T/GM$. This leads to equations of motion of the form
$$
\frac{dr}{dt} = \left( \frac{A}{B} \right)^{1/2} \, \frac{\partial \, \hat H_{\rm EOB}}{\partial \, p_{r_*}} \, ,
$$
$$
\frac{d p_{r_*}}{dt} = - \left( \frac{A}{B} \right)^{1/2} \, \frac{\partial \, \hat H_{\rm EOB}}{\partial \, r} \, ,
$$
$$
\Omega \equiv \frac{d \varphi}{dt} = \frac{\partial \, \hat H_{\rm EOB}}{\partial \, p_{\varphi}} \, ,
$$
\begin{equation}
\label{eqn12}
\frac{d p_{\varphi}}{dt} = \hat{\mathcal F}_{\varphi} \, .
\end{equation}
As explained above the EOB metric function $A(r)$ is defined by Pad\'e resumming the Taylor-expanded result (\ref{eqn4}) obtained from the matching between the real and effective energy levels (one uses a similar Pad\'e resumming for $B(r)$, or rather for $D(r) \equiv A(r) \, B(r)$). One similarly needs to resum the $\varphi$ component of the radiation reaction which has been introduced on the r.h.s. of the last equation (\ref{eqn12}). During the quasi-circular inspiral $\hat{\mathcal F}_{\varphi}$ is known (from the PN work mentioned in Section~2 above) in the form of a Taylor expansion of the form
\begin{equation}
\label{eqn13}
\hat{\mathcal F}_{\varphi}^{\rm Taylor} = -\frac{32}{5} \, \nu \, \Omega^5 \, r_{\omega}^4 \, \hat F^{\rm Taylor} (v_{\varphi}) \, ,
\end{equation}
where $v_{\varphi} \equiv \Omega \, r_{\omega}$, $r_{\omega} \equiv r \, \psi^{1/3}$ (with $\psi$ defined as in Eq.~(22) of \cite{Damour:2006tr}), and
\begin{eqnarray}
\label{eqn14}
\hat F^{\rm Taylor} (v) &= &1 + A_2 (\nu) \, v^2 + A_3 (\nu) \, v^3 + A_4 (\nu) \, v^4 + A_5 (\nu) \, v^5 \nonumber \\
&&+ \, A_6 (\nu , \log v) \, v^6 + A_7 (\nu) \, v^7 + A_8 (\nu = 0 , \log v) \, v^8 \, ,
\end{eqnarray}
where we have added to the known 3.5PN-accurate comparable-mass result the small-mass-ratio 4PN contribution \cite{gr-qc/9405062}.

\smallskip

Following \cite{Damour:1997ub}, one resums $\hat F^{\rm Taylor}$ by using the following Pad\'e resummation approach. First, one chooses a certain number $v_{\rm pole}$ which is intended to represent the value of $v_{\varphi}$ at which the exact angular momentum flux would become infinite if one were to formally analytically continue $\hat{\mathcal F}_{\varphi}$ along {\it unstable} circular orbits below the Last Stable Orbit (LSO): then, given $v_{\rm pole}$, one defines the resummed $\hat F (v_{\varphi})$ as
\begin{equation}
\label{eqn15}
\hat F^{\rm resummed} (v_{\varphi}) = \left( 1 - \frac{v_{\varphi}}{v_{\rm pole}} \right)^{-1} P_4^4 \left[ \left( 1 - \frac{v_{\varphi}}{v_{\rm pole}} \right) \hat F^{\rm Taylor} (v_{\varphi}) \right] \, ,
\end{equation} where $P_4^4$ denotes a $(4,4)$ Pad\'e approximant. It has recently been shown \cite{arXiv:0711.2628} that the {\it flexibility} in the choice of $v_{\rm pole}$ could be advantageously used to get a close agreement with NR data. The quality of the agreement with NR data obtained by such a `flexed' Pad\'e resummation is illustrated (in the small mass-ratio limit) in Fig.~2. 

%
%
\begin{figure}[h]
\centering
\includegraphics[height=8cm]{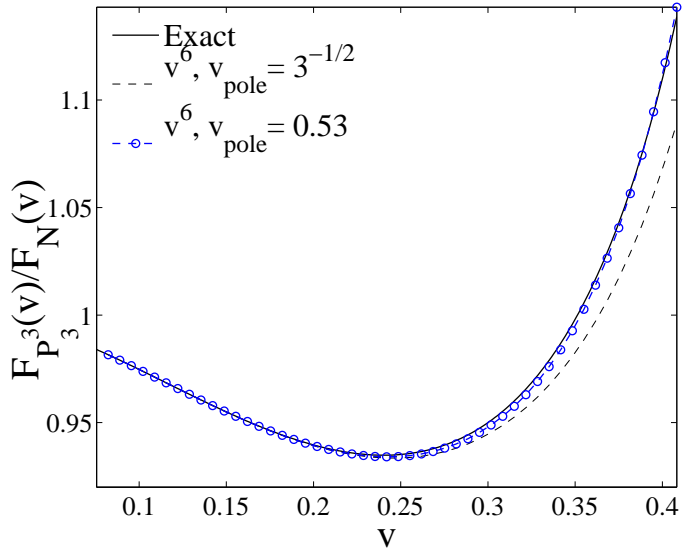}
%
%
\caption{This figure illustrates how the Pad\'e resummation of the (Newton-normalized) 3PN-expanded angular momentum flux $\hat F^{\rm Taylor} (v) = F(v) / F_N (v)$ nicely agrees with the exact, numerical flux when using a suitably `flexed' value of $v_{\rm pole}$. The solid curve is the exact, numerical flux \cite{Poisson:1995vs}. The empty circles denote the $(3,3)$ Pad\'e resummation using $v_{\rm pole} = 0.53$. The dashed curve, below the other two, denote the (3,3) Pad\'e resummation using the value $v_{\rm pole} = 1/\sqrt 3$, as originally suggested \cite{Damour:1997ub}.}
\label{F_vpole_v6.eps}       
\end{figure}

Let us note that the EOB method is not restricted to planar motions. In particular, the entire method has been extended to the case of circularized motions of {\it spinning} black holes \cite{Damour:2001tu,Buonanno:2005xu,DJS08}. In this case, one must work with more phase space variables, ${\bm q}$, ${\bm p}$, ${\bm S}_1$ and ${\bm S}_2$, and include spin effects in the radiation reaction.

\section{Effective One Body waveforms}\label{sec:5}

To end this brief review, let us sketch the definition of accurate waveforms within the EOB approach. The construction of EOB waveforms is based on several separate inputs:
\begin{itemize}
\item[$\bullet$] high-accuracy, PN-expanded inspiralling waveforms as basic inputs (see references above);
\item[$\bullet$] a specific resummation of the inspiral-plus-plunge waveform $h^{\rm insplunge} (t)$ (which includes the resummation of an infinite number of  `leading logarithms' entering the Multipolar-post-Minkowskian expansion of tail effects, the resummation of the 5PN-accurate\footnote{The 3PN-accurate part of the `non-tail' waveform is known in the comparable-mass case $\nu \ne 0$. The 4PN and 5PN pieces are only known in the test-mass limit $\nu \to 0$ \cite{gr-qc/9405062,gr-qc/9701050}.} `non-tail' $\ell = 2$, $m=2$ waveform\footnote{See \cite{arXiv:0710.0614} for an independent derivation of the non-resummed 3PN-accurate $\ell = 2$, $m=2$ waveform.}, and the parametrization of some non-quasi-circular effects) \cite{arXiv:0705.2519,arXiv:0711.2628}:
\begin{equation}
\label{eqn16}
\left( \frac{R c^2}{GM} \right) h_{22}^{\rm insplunge} (t) = -8 \, \sqrt{\frac{\pi}{5}} \ \nu (r_{\omega} \, \Omega)^2 \, F_{22}\, f_{22}^{\rm NQC} \, e^{-2i\Phi} \, ;
\end{equation}
\item[$\bullet$] a simplified representation of the transition between plunge and ring-down by smoothly {\it matching}, on a $(2p+1)$-toothed ``comb'' $(t_m - p\delta , \ldots , t_m - \delta , t_m , t_m + \delta , \ldots , t_m + p\delta)$ centered around a matching time $t_m$, the inspiral-plus-plunge waveform to a ring-down waveform, made of the superposition of several\footnote{Refs.~\cite{arXiv:0705.2519,arXiv:0712.3003} use $p=2$, {\it i.e.} a 5-teethed comb, and, correspondingly, 5 positive-frequency Kerr Quasi-Normal Modes.} quasi-normal-mode complex frequencies,
\begin{equation}
\label{eqn17}
\left( \frac{R c^2}{GM} \right) h_{22}^{\rm ringdown} (t) = \sum_N C_N^+ \, e^{-\sigma_N^+ (t-t_m)} \, ,
\end{equation}
with $\sigma_N^+ = \alpha_N + i \, \omega_N$, and where the label $N$ refers to indices $(\ell , \ell' , m , n)$, with $(\ell , m) = (2,2)$ being the Schwarzschild-background multipolarity of the considered (metric) waveform $h_{\ell m}$, with $n=0,1,2\ldots$ being the `overtone number' of the considered Kerr-background Quasi-Normal-Mode, and $\ell'$ the degree of its associated spheroidal harmonics $S_{\ell ' m} (a \sigma , \theta)$. As discussed in \cite{Buonanno:2000ef} and \cite{arXiv:0705.2519} the physics of the transition between plunge and ring-down (which was first understood in the classic work of Davis, Ruffini and Tiomno \cite{Davis:1972ud}) suggests to choose as matching time $t_m$, in the comparable-mass case, the EOB time when the EOB orbital frequency $\Omega (t)$ reaches its {\it maximum} value.
\end{itemize}

\medskip

Here, we have been describing the $3^{+2}$-PN-accurate {\it resummed} EOB waveform introduced in \cite{arXiv:0705.2519,arXiv:0711.2628}. A less accurate, {\it restricted} ($0$-PN) EOB waveform (together with a simplified matching procedure; $\delta \to 0$) has been used by Buonanno and collaborators \cite{arXiv:0706.3732}.

\smallskip

Finally, one defines a complete, quasi-analytical EOB waveform (covering the full process from inspiral to ring-down) as:
\begin{equation}
\label{eqn18}
h_{22}^{\rm EOB} (t) = \theta (t_m - t) \, h_{22}^{\rm insplunge} (t) + \theta (t-t_m) \, h_{22}^{\rm ringdown} (t) \, ,
\end{equation}
where $\theta (t)$ denotes Heaviside's step function.

\smallskip

An example of this complete EOB waveform is represented in Fig.~3. 

%
%
\begin{figure}[h]
\centering
\includegraphics[height=8cm]{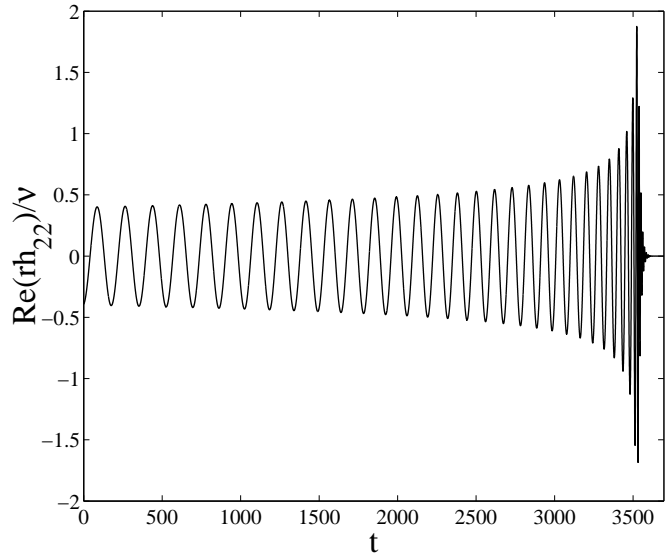}
%
%
\caption{This figure illustrates a complete resummed EOB quadrupolar ($\ell = 2$, $m=2$) waveform (\ref{eqn18}) (using $a_5 = 20$, $\bar a_{RR} = 30$ and a suitably flexed $v_{\rm pole}$), with, about, 29 GW cycles of inspiral, $\sim 1$ GW cycle during plunge, and $\sim 4$ GW cycles of ring-down. Ref.~\cite{arXiv:0711.2628} has shown that the inspiral part of the EOB waveform agreed with the Caltech-Cornell NR data \cite{arXiv:0710.0158} within $\pm 0.001$ GW cycles, while Ref.~\cite{arXiv:0712.3003} has shown that the late-inspiral-plunge-ringdown part of the EOB waveform agrees with an Albert Einstein Institute NR waveform within $\pm 0.005$ GW cycles.}
\label{h22_by_nu_eob.eps}       
\end{figure}

Recent comparisons between EOB waveforms and full, 3-dimensional NR waveforms have exhibited an excellent level of agreement. After a preliminary comparison done in \cite{gr-qc/0610122}, Buonanno et al. \cite{arXiv:0706.3732} compared {\it restricted} EOB waveforms to NR waveforms computed by the NASA-Goddard group. In the equal-mass case ($\nu = 1/4$), they found that the dephasing between (restricted) EOB and NR waveforms (covering late inspiral, merger and ring-down) stayed within $\pm 0.030$ GW cycles over 14 GW cycles. In the case of a mass ratio $4:1$ ($\nu = 0.16$), the dephasing stayed within $\pm 0.035$ GW cycles over 9 GW cycles. The {\it resummed} EOB waveform has been compared to two different (equal-mass) NR waveforms: (i) in the comparison with the very accurate inspiralling simulation of the Caltech-Cornell group \cite{arXiv:0710.0158} the dephasing stayed smaller than $\pm 0.001$ GW cycles over 30 GW cycles (and the amplitudes agreed at the $\sim 10^{-3}$ level) \cite{arXiv:0711.2628}; (ii) in the comparison with a late-inspiral-merger-ringdown NR waveform computed by the Albert Einstein Institute group, the dephasing stayed smaller than $\pm 0.005$ GW cycles over 12 GW cycles. See Ref.~\cite{DN08} for another EOB/NR comparison. All these comparisons have used the natural {\it flexibility} of the EOB formalism \cite{Damour:2001tu,gr-qc/0211041}, {\it i.e.} the possibility to tune EOB parameters representing yet uncalculated effects (such as $a_5$ in $A(r)$ or $v_{\rm pole}$).

\smallskip

We hope to have shown here the usefulness of the EOB formalism in accurately describing the general relativistic motion and radiation of coalescing black hole binaries.

\vglue 1cm

\noindent {\bf Acknowledgments.} I thank Alessandro Nagar for many fruitful interactions, and the conveners of the Second Stueckelberg meeting for organizing a stimulating workshop. I am also grateful to Alessandro Nagar and Marie-Claude Vergne for help with the figures, and Eric Poisson for providing the data of Fig.~2.

\newpage

%



\end{document}